# ProtoSound: A Personalized and Scalable Sound Recognition System for Deaf and Hard of Hearing Users

Dhruv Jain

University of Washington, Seattle, United States; Google, Mountain View, United States, djain@cs.uw.edu

Khoa Huynh Anh Nguyen

University of Washington, Seattle, United States, akhoa99@cs.uw.edu

Steven Goodman

University of Washington, Seattle, United States, smgoodmn@uw.edu

Rachel Grossman-Kahn

University of Washington, Seattle, United States, rachel.illowsky@gmail.com

Hung Ngo

University of Washington, Seattle, United States, hvn297@cs.uw.edu

Aditya Kusupati

University of Washington, Seattle, United States, kusupati@cs.washington.edu

Ruofei Du

Google Research, San Francisco, CA, United States, ruofei@google.com

Alex Olwal

Google Research, Mountain View, CA, United States, olwal@acm.org

Leah Findlater

University of Washington, Seattle, United States, leahkf@uw.edu

Jon E. Froehlich

University of Washington, Seattle, United States, jonf@cs.uw.edu

Recent advances have enabled automatic sound recognition systems for deaf and hard of hearing (DHH) users on mobile devices. However, these tools use pre-trained, generic sound recognition models, which do not meet the diverse needs of DHH users. We introduce *ProtoSound*, an interactive system for customizing sound recognition models by recording a few examples, thereby enabling personalized and fine-grained categories. ProtoSound is motivated by prior work examining sound awareness needs of DHH people and by a survey we conducted with 472 DHH participants. To evaluate ProtoSound, we characterized performance on two real-world sound datasets, showing significant improvement over state-of-the-art (*e.g.,* +9.7% accuracy on the first dataset). We then deployed ProtoSound's end-user training and real-time recognition through a mobile application and recruited 19 hearing participants who listened to the real-world sounds and rated the accuracy across 56 locations (*e.g.,* homes, restaurants, parks). Results show that ProtoSound personalized the model on-device in real-time and accurately learned sounds across diverse acoustic contexts. We close by discussing open challenges in personalizable sound recognition, including the need for better recording interfaces and algorithmic improvements.

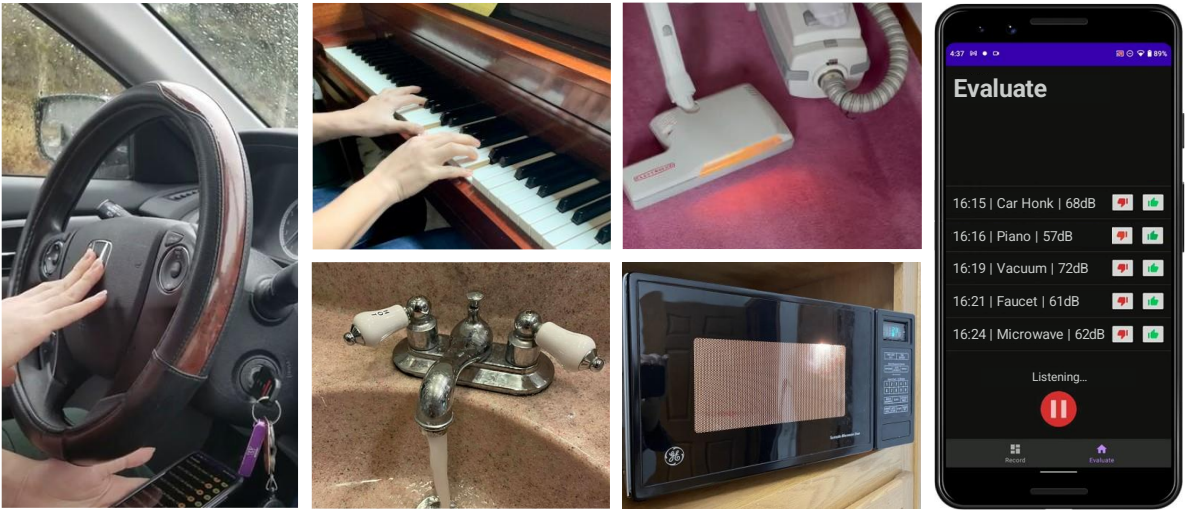

Figure 1: ProtoSound is a technique to customize a sound recognition model using very few recordings, enabling the model to scale across contextual variations of sound (*e.g.,* water flowing on a stainless steel *vs.* a porcelain sink) and support new user-specific sound classes (*e.g.,* a piano). Images show some example sound categories that were trained and recognized during our field evaluation using an experimental mobile app built off ProtoSound. See our supplementary video for details.

## 1 INTRODUCTION

Sound recognition can provide important information about the environment, human activity, and situational cues to people who are d/Deaf and hard of hearing (DHH) [5,14,23]. Recent advances in machine learning and signal processing have enabled automatic sound recognition—a feature now available on both major mobile platforms: Google Android [74] and Apple iOS [43]. However, prior sound recognition systems for DHH users [25,26,33] use generic models that are trained on large sound corpora and do not support end-user personalization—such as training on new sound categories (*e.g.,* a new custom home appliance) or a specific sound (*e.g.,* my child's voice or pet's dog bark) [5,14].

In this paper, we present *ProtoSound*, an interactive system that allows users to personalize a sound recognition engine by recording custom sounds (Figure 1). Unlike traditional data-intensive machine learning approaches, users can customize a model using only a few sample recordings (*e.g.,* five for each sound). While prior machine learning work (*e.g.,* [58,66]) has performed algorithmic experiments of "*few-shot*" sound recognition, we contribute the first useable system by integrating user-centric features such as: (1) on-the-fly training for difficult-to-produce sounds (*e.g.,* fire alarms, sirens) and (2) handling contextual soundscape variations (*e.g.,* homes *vs.* outdoors). In contrast, traditional few-shot approaches require the full training set to be available beforehand [15,60] and do not generalize well across contexts [8].

To guide ProtoSound's evaluation, we conducted a large-scale survey with 472 DHH participants, which uncovered key personalization preferences such as the minimum number of custom sounds to support and the maximum desired recording effort. We then used these insights to design three experiments: quantitative evaluations on two real-world datasets and a field study. On a dataset of sounds recorded by hearing people in multiple contexts, ProtoSound outperformed the best baseline model by a 9.7% accuracy margin (88.9% vs. 79.2%). The average accuracy (88.9%) was close to the ground truth obtained by manual human labeling (91.3%). On an additional dataset of sounds recorded by DHH people in and around their homes, ProtoSound's average accuracy was 90.4%. In comparison, the dataset's label accuracy rated by a hearing person was 94.5%.

While the above results are promising, they do not reflect an actual system use. Thus, we deployed ProtoSound's end-user training and real-time recognition through a mobile application and conducted a field evaluation with 19 hearing participants—to our knowledge, the first evaluation of few-shot sound recognition in the field. While our ultimate goal is a long-term study with DHH users, demonstrating real-world efficacy and improving the system is an important step before deployments with the target population; hence, we recruited hearing users who could reliably listen to the real-world sounds and evaluate ProtoSound's recognition accuracy. Results show that ProtoSound trained the model on-device through low end-user effort and accurately learned sounds in a range of acoustic environments (*e.g.,* homes, restaurants, grocery store, parks, and streets). However, errors arose due to recording mistakes (*e.g.,* incorrect labels, overlapping sounds), pointing to a need to develop better user interfaces in the future.

In summary, our work contributes: (1) a real-time, personalized sound recognition system for DHH users, (2) results from a suite of evaluation experiments providing insight into the feasibility of few-shot sound recognition in the field, (3) findings from a large-scale survey identifying personalization preferences of 472 DHH participants, and (4) two open-source artifacts: a Python-based implementation of the ProtoSound pipeline deployable to any device, and an Android-specific on-device implementation.

## 2 BACKGROUND AND RELATED WORK

We provide background on and contextualize our work within sound awareness needs and systems for DHH users, acoustic signal processing algorithms, and relevant machine learning approaches.

### 2.1 Sound Awareness Needs of DHH users

ProtoSound is informed by the diverse sound awareness needs of the DHH community. A person belonging to the DHH community may identify as Deaf (capital 'D'), deaf (small 'd'), or hard of hearing [6,73]. Individuals who identify as Deaf follow an established set of norms, behaviors, and language (called 'Deaf culture' [6,32,45]). In contrast, deaf or hard of hearing individuals connect to deafness audiologically and refrain from membership to a particular community [6,45]. These individuals do not have a distinct cultural identify of their own and may choose to interact with either Deaf or hearing people based on comfort. These cultural differences may influence sound awareness preferences. For example, two large-scale surveys with DHH participants [5,14] found that hard of hearing users may be more interested in some sounds (*e.g.,* phone ringing, speech) than d/Deaf users.

While accounting for diversity of preferences, prior work also highlights several general sound awareness needs among DHH people [5,14,24,40]. For example, within the sound characteristics such as volume or duration, sound identity is the most desired, with all DHH sub-groups generally ranking urgent sounds (*e.g.,* fire alarm, siren) as most important, followed by sounds indicating human activities (*e.g.,* doorbell, footsteps) and appliance alerts (*e.g.,* microwave beep, kettle boiling) [5,14]. Additionally, the relevance of sound information may vary with social contexts (*e.g.,* family *vs.* strangers) [14,25] and physical locations (*e.g.,* at home *vs.* while mobile) [5,14,20]. This points to the need for the DHH users to be able to personalize their sound awareness systems.

### 2.2 Sound Awareness Systems for DHH users

Commonly used technologies by DHH users include flashing doorbells and vibratory wake-up alarms that provide visual and haptic alternatives to specific auditory information. While useful for their specific applications, these devices do not offer a general alternative to environmental sounds.

In pioneering work, Matthews *et al.* [40] built a desktop-based prototype that used sound visualizations (*e.g.,* spectrograph, rings) to convey basic sound information (*e.g.,* pitch, source location) in an office setting. The same team later developed a Personal Digital Assistant (PDA) app for DHH users to request human-assisted transcription of speech and non-speech sounds in the last 30 seconds of audio [39]. More recent work aimed to provide broader sound recognition support with pre-trained deep-learning models [25,26,59]. For example, Sicong *et al.* [59] leveraged convolutional neural networks (CNNs) to build and evaluate a smartphone-based app that sensed and classified nine environmental sounds (*e.g.,* door knock, bell ringing). Jain *et al.* [25] conducted field deployments of a smarthome sound awareness system that recognizes 19 sounds (*e.g.,* microwave beeps, water running) in the homes of DHH users. Participants found the system useful for knowing about home activities but expressed a desire to personalize the system to sounds specific to their homes (*e.g.,* children and pets).

In terms of personalizable systems, Bragg *et al.* [5] developed a mobile app to recognize sounds that were recorded by DHH participants in a user study. However, this preliminary Gaussian Mixture Model (GMM)-based approach classified only two sounds in an office setting with limited accuracy, and is unlikely to represent varied use cases, sound, and environmental noise in the daily life of DHH users. Jain *et al.* [26] built a smartwatch-based sound recognition app that allowed DHH end-users to filter notifications for undesired sounds. This sound filtering, however, was performed on the interface after prediction from a generic model and did not support adding or modifying sound classes through user-provided recordings. In a user evaluation, participants found the app useful in general, but less accurate in noisy environments, and wanted to add custom sounds (*e.g.,* footsteps).

We build on the work above by examining a personalized sound recognition system that can support custom sounds in a diversity of contexts.

### 2.3 Acoustic Signal Processing Algorithms

Acoustic signal processing involves extracting meaningful information from the time or frequency domain of an audio signal [7,36]. Easy-to-compute information such as zero-crossing rate (ZCR), short-time energy (STE), and spectral flux (SF) [12,44,49,52] perform reasonably well on clean sound files, but fail to account for real-world acoustic variations [36]. Thus, autoregression-based features (*e.g.,* Linear Prediction Coefficients (LPCs) [38]) were developed to capture variations in speech and music. For sound recognition specifically, cepstral features that model the human auditory system such as Mel-frequency cepstral coefficients (MFCCs) or Harmonic Cepstral Coefficient (HCC) [7,11] are common. For example, in Bragg *et al.*'s work detailed above [5], MFCC features were fed to a GMM-based classifier. Lu *et al.* [35] trained a two-step classifier and used ZCR and SF features to distinguish pre-defined sound events (music, speech, and ambient noise) and MFCCs to identify new sounds.

The above stationary features, while providing a good representation of psychoacoustic properties (*e.g.,* loudness, pitch, timbre), do not model the temporal variation in real-world sounds. Hence, non-stationary methods based on wavelets [10], sparse-representations [9], or power-spectrum [29] are often used in combination with stationary features to encode temporal variation on a frequency spectrum. Of these, deep-learning architectures have shown the most promise by modeling the subtleties and non-linearities in acoustic data, distinguishing a large variety of real-life sound events. ProtoSound uses a lightweight CNN model commonly used for image classification on mobile devices [57], but fine-tuned on online sound effect libraries.

### 2.4 Relevant Machine Learning Training Paradigms

Traditional supervised training paradigms are useful for specific tasks such as gunshot detection [16] or intruder alerts [3] but require a large amount of in-situ data to work in diverse contexts [15]. For more modest training set sizes, relevant machine learning approaches include transfer learning [63], a supervised training method that uses limited training examples to fine-tune a model previously trained on large datasets from a different domain (*e.g.,* image classification). Likewise, co-training approaches [47,64] use a small number of examples and a large unlabeled set to create a model with better classification performance. Our system uses meta learning [65], a learning approach that allows models to recognize previously unseen classes or adapt to new environments with very few labelled training instances. This approach has recently been explored in many domains, including computer vision [15,53], acoustic event detection [58,66], and natural language processing [50,72].

Most similar to our work is *ListenLearner* [68], which provides a platform for learning new classes through one-shot user labelling. ListenLearner starts with a pre-defined set of classes that it uses to learn representations of new sounds by recording a large number of samples and prompting the user for labelling (*e.g.,* "*what sound was that?*"). However, this semi-supervised approach requires longitudinal deployments for recording many samples, while our approach allows for quicker adaptation to new environments through fewer training examples. Furthermore, by prompting users for feedback at unspecified times, ListenLearner assumes that users have domain knowledge (*i.e.*, they can listen to and identify a recently occurring sound)—an assumption that may not hold for DHH users. Our intentional recording approach may enable users to leverage visual and contextual cues for recording (*e.g.,* by seeing that a faucet is turned on). Finally, unlike ProtoSound, ListenLearner does not support customizing existing classes (*e.g.,* my dog *vs.* a generic dog). Acoustic distribution of sound classes may vary widely across acoustic contexts [37] and using existing class representations may not generalize well.

## 3 THE PROTOSOUND SYSTEM

ProtoSound is an interactive system for personalizing a sound recognition model in real-time using few custom recordings. ProtoSound uses prototypical networks [60], one of the most efficient algorithms for few-shot classification and extends the traditional training pipeline to incorporate user-centric features for real-world deployment—such as a technique to accommodate varying contexts of use, and a library of difficult-to-produce sounds preferred by DHH people. Throughout the design of ProtoSound, we worked with individuals of the DHH community, including the lead author who is DHH, and a co-author, who is an ASL interpreter.

### 3.1 System Design

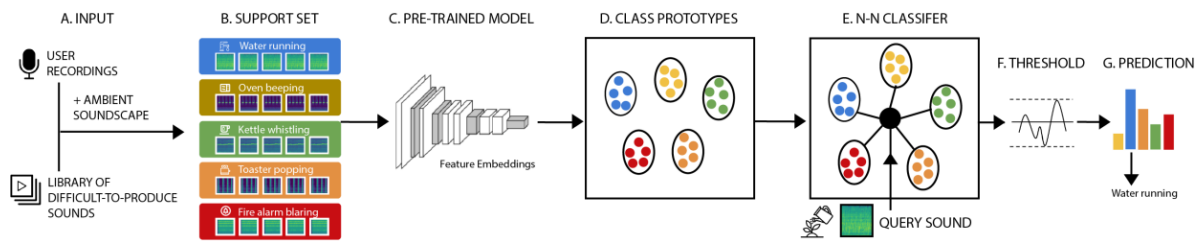

Figure 2: Our ProtoSound few-shot sound recognition pipeline. For the desired sound classes, users either record a few samples of sounds on their own or select from our online library of difficult-to-produce sounds (*e.g.,* for fire alarms, sirens) (A). To accommodate contextual shifts, these sound samples are then mixed with the ambient soundspace of a place, leading to the *support set* (B). This support set is then fed into our pre-trained sound recognition model (C) to generate *class prototypes*, which is the representation of each sound class in the feature space (D). During prediction, a query sound is then compared with these stored class prototypes using a nearest neighbor distance metric (*i.e.,* N-N classifier) (E). When our confidence threshold is passed (F), the nearest class is outputted as the prediction (G).

ProtoSound's sound-sensing pipeline involves two phases: model personalization and prediction. Model personalization includes personalizing the model from a set of user recordings (Figure 2B). During this phase, *log-mel spectrograms* [22] of user recordings in a context, or samples from our library of difficult-to-produce sounds are fed into the model to extract feature embeddings (Figure 2C) We use log-mel spectrogram input features since they have historically shown better performance than other alternatives (*e.g.,* MFCCs) with CNN architectures [34]. The extracted embeddings for each class are averaged, resulting in class prototypes, which are representations of each class in feature space (Figure 2D). These class prototypes are used for predicting a new sound using a nearest-neighbor classifier—that is, we output the class nearest to a query sample by calculating the Euclidean distance in the feature space (Figure 2E—G). In addition, to aid real-world use, ProtoSound contains several user-centric features: context generalization, a library of difficult-to-produce sounds, and open-set classification.

*3.1.1 Context Generalization*

Ideally, the users should record sounds for model personalization in each context. However, in real-life, users may move across auditory contexts (*e.g.,* inside homes to outdoors), and may reuse a model trained in one context in another—for example, a model trained on sounds such as water running in the home could also be used outdoors. Such context shift often introduces novel acoustics conditions (*e.g.,* background noise, changing data distributions) and a model may not generalize well. This is particularly an issue with meta-learning approaches which tend to overfit the model on context specific data [58]. Cross-setting generalization methods increase the robustness of classification algorithms by adapting them to a target context [19].

ProtoSound uses a custom, data-driven cross-setting generalization technique [19]: augmenting the samples procured from a source domain that the model was previously trained on (*e.g.,* a home) with the ambient soundscape of the target domain (*e.g.,* an outdoor location), using the following equation:

$$\text{Target sample} = (1 - \alpha) * \text{source sample} + \alpha * (\text{target soundscape} - \text{source soundscape})$$

To determine $\alpha$, we performed iterative experiments on two benchmark sound datasets (*ESC-50* [51] and *UrbanSound8k* [56]) and selected an optimum value of 0.3. Although we chose a single $\alpha$ value to reduce the number of tunable parameters, it can be set to change based on a particular auditory context shift (*e.g.,* homes-to-outdoors may have a different value than outdoors-to-restaurant). Note that while the soundscape may vary across samples recorded from one context, our context generalization scheme only needs an estimate of the background noise to determine the bounds of a feature space in a context. Beyond accommodating context shifts, cross-setting generalization could also help mitigate other possible acoustic variations, such as those caused by different recording devices (*e.g.,* a model created on a laptop may be used on a phone).

*3.1.2 Existing Library of Difficult-to-Produce Sounds*

ProtoSound requires sound recordings for personalization. However, in real-life, there may be sounds that are highly desired by DHH users but do not occur spontaneously for recording (*e.g.,* fire alarms, sirens). To support training for these difficult-to-produce sounds, ProtoSound contains samples of 10 sound categories preferred by DHH people [5,26] (*e.g.,* fire/smoke alarms, babies crying, sirens, bird chirps), procured from a high-quality online library, *FreeSound*. These sound samples were manually cleaned (removing noise, deleting silences) by three hearing authors and are available in ProtoSound's repository. During training, these sounds are augmented with the ambient soundscape of the target domain.

*3.1.3 Open Set Classification*

Most sound classification tools assume a closed-set classification scenario, with a fixed set of predefined classes to distinguish [1]. In real-world, however, the underlying data distributions of soundscapes are often unknown and can change over time with new classes becoming relevant [1,27]. To accommodate this issue, researchers have introduced open-set classification approaches (*e.g.,* [42]), where an algorithm can also classify a given sound as "unknown". ProtoSound uses the following open-set classification algorithm adapted from Júnior *et al.* [27]:

Let $d_1$ and $d_2$ be the respective Euclidean distance of a query sample from the nearest and the second nearest class prototype in feature space. Then, we calculate the ratio:

$$R = \frac{d_1}{d_2}$$

If $R$ is less than or equal to a specified threshold $T$, the query sample is classified as the same label as the nearest class. Otherwise, it is ignored. Following our internal experiments, we used a $T$ value of $0.6$.

In addition to the above algorithm, ProtoSound uses an end-user tunable loudness threshold (default value: 45dB, equivalent to an AC hum). During prediction, any query with average loudness below this value is ignored.

**3.2 System Implementation**

*3.2.1 Model Architecture and Pre-training*

We implemented ProtoSound using a MobileNetV2 architecture [54]—a state-of-art CNN for mobile devices, measuring about 8MB. Past few-shot learning work (e.g., [55]) did not find improvements from using bigger networks like ResNets [60] due to the risk of overfitting on sparse data [55,59]. We pre-trained the model using a train set compiled from six online sound effect libraries—*Freesound* [17], *BBC* [75], *Network Sound* [76], *UPC* [77], *TUT* [41] and *TAU* [2]—each of which provide a collection of high-quality, pre-labeled sounds. We selected sound categories for which we found more than 1000 clips, which included a total of 35 common sounds from different contexts (*e.g.,* homes, urban, outdoors, see Table 1). Clips were downloaded, converted to a single format (16KHz, 16-bit, mono), and silences greater than one second were removed, resulting in 38.8 hours of sound data.

We segmented each clip into one second audio segments and computed short-time Fourier Transforms using a 25ms sliding window and 10ms step size (frequency range from 20Hz to 8000Hz), which yielded a 96-length spectrogram. We then converted our linear spectrogram into a 64-bin log-scaled Mel spectrogram and generated a 100 X 64 input frame for every one second of audio. To these log-mel spectrograms, we applied Cepstral Mean and Variance Normalization (CMVN) [61] before inputting into the model. For training, we used a cross entropy loss function with an Adam optimizer [30].

*3.2.2 Selection of the Prototypical Networks Algorithm*

We selected prototypical networks as our base algorithm following our performance comparison experiment with five state-of-the-art few-shot learning approaches: MAML [15], FoMAML [15], Reptile [46], ANIL [54] and Prototypical Networks [60]. For our experiments, we used three benchmark sound datasets: *AudioSet* [18], *ESC-50* [51], and *UrbanSound8k* [56]. Results reflect past work [58,66] with prototypical networks performing the best (*avg. accuracy*=95.6%) followed by ANIL (*avg. accuracy*= 93.4%), Reptile (*avg. accuracy*=91.7%), FoMAML (*avg. accuracy*=90.8%), and MAML (*avg. accuracy*=90.6%). Prototypical Networks also had the lowest training time.

*3.2.3 Open-Source Release*

For researchers and practitioners to build on our work, a PyTorch-based implementation of ProtoSound with our pre-trained MobileNetV2 model is available at https://github.com/makeabilitylab/ProtoSound. The code can support any number of classes and can run on any device with a Python interpreter. The sound samples can be supplied from live microphone or file input. For live prediction, our code samples the microphone at 16KHz and segments the input into 1-second buffers, which serve as query samples.

## 4 SURVEY: PERSONALIZED SOUND RECOGNITION (472 DHH PEOPLE)

While past studies have shown that personalized sound recognition is generally desired among DHH people [5,25,26], the specific customization preferences are as yet unknown (*e.g.,* how many custom sound classes are desired in a context, maximum recording effort users are willing to put). We conducted an online survey with DHH participants to better understand these preferences and to shape ProtoSound's evaluation (*e.g.,* the number of classes to use in our experiments).

### 4.1 Participants

We used Google surveys [78], which targets users of the Google Opinion Rewards Android app [79]. Due to our institutional policy, we could not ask about identity (*e.g.,* deaf vs. Deaf) or hearing loss levels. Instead, we relied on a DHH assistive technology screener, and targeted respondents who indicated use of "TDD, TTY, or closed captions" (58% of the selected 472 participants), "Hearing aid" (19%), "Real-time captions (*e.g.,* CART)" (29%), "Android Live Transcribe & Sound Notifications" (18%), and/or "Other hearing assistive devices" (9%) in a survey question. 511 respondents satisfied this criterion, but we excluded 39 who misunderstood our survey (*e.g.,* confused sound events with calendar notifications) or provided invalid responses. The remaining 472 participants were adults (18 and older) across US states and territories with 55% men, 43% women, and 2% of unknown gender. All participants used Android smartphones and were compensated up to $1 USD.

### 4.2 Survey Design

The 10-question survey took about 3 minutes to complete (*avg.*=2 min 47 s, *SD*=3 min 52 s) and asked about the use of the current Android sound recognition feature [74], its usefulness, interest in recording sounds for a future personalized system, and the number of sounds a personalized system should support. For the complete survey questionnaire, see Appendix A1.

### 4.3 Results

About 34% (162) of the 472 participants used Android sound recognition multiple times a week (22% used it daily). Of these weekly/daily users, 89% rated it useful (40%: extremely useful). Among the remaining 310 participants, the majority (71%) rated its usefulness as neutral and 24% indicated they were not aware of the feature.

Participants were also able to select from a list of options for what prevented them from using sound recognition more frequently. Among the weekly/daily users, 81% were concerned about system accuracy (47%: too many notifications, 17%: incorrectly recognized sounds, 22%: missed sounds, 18%: false alerts) and 33% felt that the recognition was too generic (21%: *"might not recognize some sounds I care about"*, 15%: *"can't select the sounds I want"*), pointing to a need for personalization. Indeed, 73% of the weekly/daily users indicated that they would be interested in recording sounds to personalize the system.

When asked to select the minimal number of sounds a sound recognition technology needs to support in each context (*e.g.,* kitchen, bedroom, restaurant) to be useful, a majority (74%) selected 6 sounds or less (35%: 1-3, 39%: 4-6), indicating that a few medium-to-high priority sounds are desired in a location.

In two open-ended questions, participants specified how much effort (number of sounds and time) they were willing to spend on recording their personal sounds in each context (*e.g.,* kitchen, bedroom). 71% were not willing to record more than 15 sounds and wanted to spend less than 25 minutes for each context.

### 4.4 Discussion

Our findings suggest that nearly a third of our DHH participants use sound recognition multiple times a week and most of those users find it useful (89%). Our results also suggest that ProtoSound could increase usage and value with personalized models, and help users become aware of sounds that are specific to them or their environment. The majority (74%) of our participants indicated that 6 sounds or less could suffice in each context and expressed a desire to not spend significant recording time or effort. These findings suggest that a few medium-to-high priority sounds could cover the needs of a majority of DHH users, and that a low-effort, few-shot experience is important.

## 5 EXPERIMENT 1: ON REAL-WORLD SOUNDS COLLECTED BY HEARING RESEARCHERS

Our first experiment evaluated ProtoSound on sounds recorded by hearing researchers in real-world settings.

### 5.1 Experimental Setup

**Test Dataset:** Since commonly used 'synthetic' sound classification benchmarks (*e.g.,* ESC-50 [51], UrbanSound8k [56]) do not mimic the real-world conditions (*e.g.,* background noise, overlapping sounds), we created a 'naturalistic' test set by compiling datasets of real-life sound recordings from two prior HCI works [25,26]. It contains samples for 22 common sounds preferred by DHH people [5,24] and ambient soundscapes recorded by hearing researchers in a total of 21 locations (e.g., homes, university labs, lounges, parks, and urban streets) Thirteen sound classes also exist in the dataset used to pre-train the model; nine are new (Table 1). These recordings were converted to the same format as the train set (16KHz, mono), resulting in 4.5 hours of data.

Table 1: Sound classes in our train (online libraries) and test (real-life recordings) sets. Bolded classes appear in both sets.

| Train dataset | Test dataset |
| --- | --- |
| Fire/smoke alarm, Alarm clock, Door knock, Typing, Door open/close, Vacuum cleaner, Toothbrush, Toilet flush, Water running, Hair dryer, Wood creak, Sawing, Hammering, Drilling, Dog bark, Cat meow, Cricket, Bird chirp, Engine idling, Vehicle running, Car horn, Footsteps, Breathing, Cough, Snore, Speech, Laugh, Clap, Wind, Train, Helicopter, Aircraft, Gunshot, Glass breaking, Fireworks | **Fire/smoke alarm**, **Alarm clock**, **Door knock**, Doorbell, **Door open/close**, Microwave, Cutlery, Dishwasher, **Water running**, Kettle Whistle, Phone ringing, Washer/dryer, **Dog bark**, **Cat meow**, **Bird Chirp**, Baby crying, **Vehicle running**, **Car horn**, Siren, **Cough**, **Snore**, **Speech** |

**Tasks:** In a meta-learning paradigm, a model successfully *learns to learn* [15] on a set of few-shot tasks sampled from a labelled dataset. Each meta-learning task [15] includes a support set, containing a few examples for model training, and a query set, consisting of examples for accuracy evaluation. In algorithmic terms, a meta-learning task is defined as:

> *Given a support set of N classes (called N-way) and K-samples for each class (called K-shot, where K is small, usually < 10), the aim is to classify samples on a query set along the N classes.*

For our experiments, we used the 5-way, 5-shot setting for two reasons. First, it aligns with past evaluations of generic-model systems [5,25,26] where DHH users found 3-5 medium-to-high priority sounds per context to be sufficient. Second, a low number of classes and samples per class will reduce the user's recording time—in our survey, 74% participants desired six or fewer classes and 71% wanted to spend less than 25 minutes recording.

**Baseline Algorithms:** Beyond evaluating our ProtoSound pipeline, we also compared its performance with two baseline approaches: the traditional prototypical networks pipeline [60], which is the current state-of-the-art in few-shot classification [58,66] and a fully supervised method used in commercial systems [26,43,74]. For the supervised method, we pre-trained the model with our train set (Table 1). After pre-training, we replaced the last layer by a randomly initialized linear layer with output dimension of 5 (number of ways) and fine-tuned on the test tasks described in our experiments below.

### 5.2 Specific Experiments and Results

#### 5.2.1 Overall Accuracy

To calculate the overall accuracy, we randomly sampled 100 tasks from our real-life test set, each of batch size 100 containing 25 support samples (5 shot × 5 way) and 75 query samples (15 samples per way). After passing data through the model, we performed a clip-level prediction by aggregating the probabilities for each second of data and outputting the most likely prediction. On average, ProtoSound achieved 88.9% accuracy (*SD*=5.6%), which was significantly higher than the baselines: traditional prototypical networks achieved 79.2% (9.7% less than ProtoSound, pairwise t-test was significant: $t_{99}$=8.8, *p*<.001) and supervised fine-tuning achieved 70.6% (18.3% less, *t*=15.7, *p*<.001). Improvement over supervised fine-tuning approach is expected since, unlike them, few-shot recognition approaches tend to work well with limited data [58]. Regarding the traditional prototypical networks—the state-of-the-art in few-shot recognition—ProtoSound performed better since it can better handle the soundscape variations (*e.g.,* background noise) in each context, owning to our context generalization scheme. This is better demonstrated in the following context-specific accuracy experiment.

#### 5.2.2 Context-Specific Accuracy

Our test set contains samples from three contexts: homes (kitchen, bedroom, and living room), offices (university labs and lounge), and outdoors (parking lots, parks, and streets). As sound quality may vary across context, we also calculated the context-specific accuracies of ProtoSound and the two baselines. As expected, for all three approaches, the accuracy was higher in quiet environments of homes and offices compared to outdoors (Figure 3a). However, the accuracy difference between quiet and noisy environments (homes vs. outdoors) was much lower for ProtoSound (3.6%) than the baselines (13.8% and 16.0% respectively), suggesting that ProtoSound can better generalize across contexts. Figure 4 shows the low-dimensional projections of embeddings obtained from the three approaches in an outdoor context.

Note that we did not calculate per-class accuracies due to the limitation of the few-shot evaluation—each individual test includes a random combination of five classes from our dataset. The accuracy of each class depends heavily on which other four classes are chosen for a specific test (*e.g.*, doorbells perform poorly with phone rings), hence aggregating class performance across multiple tests is counterintuitive.

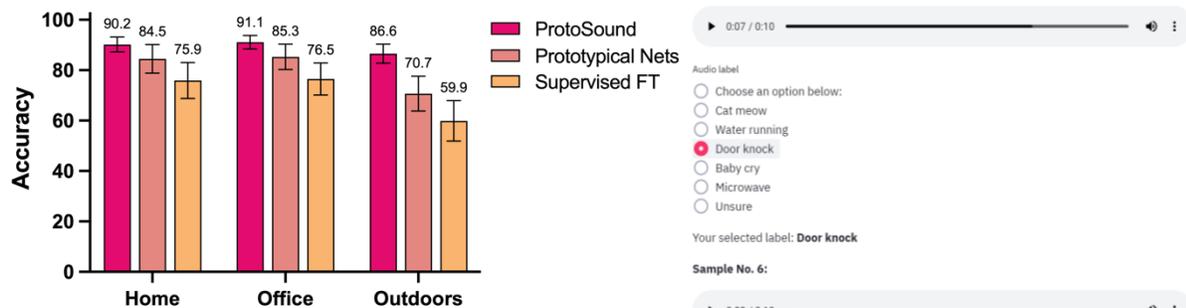

Figure 3: (a) Context-specific accuracies of ProtoSound and two state-of-the-art baseline approaches: prototypical networks and simple supervised fine-tuning. Error bars represent 95% confidence intervals. (b) Snapshot of a web-app we built to collect human labels on our real-life test set.

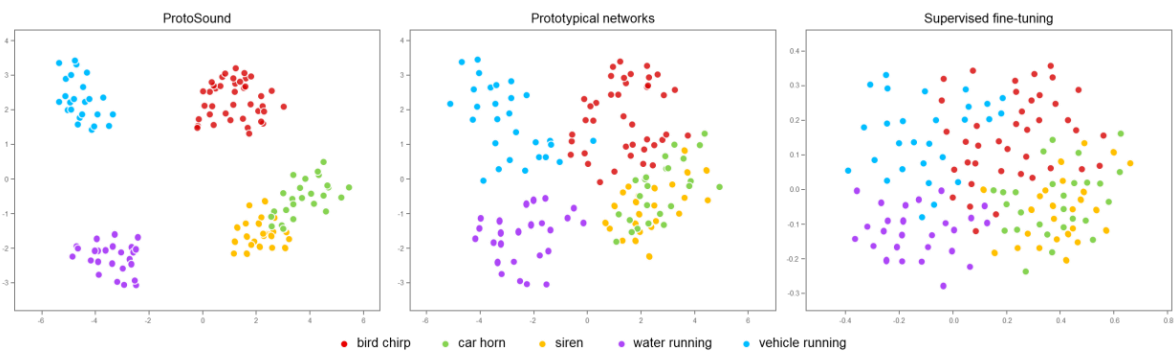

Figure 4: A visualization of t-SNE projections of embeddings obtained from ProtoSound and the two baselines: traditional prototypical networks and supervised fine-tuning for samples of five sounds in an outdoor context. Note that even for overlapping sounds "car horn" and "siren", the clusters from ProtoSound are reasonably separated in contrast to the highly overlapping clusters of the two baselines.

#### 5.2.3 Comparison to Manual Labels

To obtain ground truth performance, we recruited people to label our test set. Humans offer an excellent gold standard as they can utilize contextual knowledge from a lifetime of real-world experiences. Similar to *Ubicoustics* [33], we created a web-app (Figure 3b) that mimicked our ProtoSound model personalization and testing task. The app randomly sampled 5 sounds and 20 samples for each sound from our test set (total 100 samples) which were divided into support (25 samples) and query set (75 samples).

We then recruited six hearing participants from our research group. Each participant listened to the support samples for "learning" and categorized the samples from the query set among the 5 classes (Figure 3b). Similar to our open-set classification approach, participants could also select an "unsure" option if they thought a sample did not belong to any of the listed classes. Each participant analyzed three batches, resulting in 6 participants × 3 batches × 75 = 1350 evaluations.

Average accuracy of participants' labels was 91.3% (*SD*=4.8%). Participants revealed two factors that made it challenging to correctly classify some samples: (1) noise (*e.g.,* silence, or too much background noise) and (2) interclass similarities—that is, sounds that were very similar (*e.g.,* doorbells and phone rings). In comparison, our model achieved 92.9% average accuracy for the same setup (*SD*=4.3%), which is close to human performance; a paired t-test was not significant ($t_{17}$=0.8, *p*=0.4). On further investigation, we found that, like humans, the errors were most prominent for similar sounding events (*e.g.,* alarm clock and phone ringing), which were often confused.

# 6 EXPERIMENT 2: ON REAL-WORLD SOUNDS COLLECTED BY DHH PEOPLE

While the experiment above was necessary to contextualize ProtoSound within prior work, the dataset was collected by hearing people and could lead to representation bias [67]. To combat this, we also evaluated ProtoSound's performance on sounds recorded by DHH participants in a prior study [21].

This data was collected and labelled by 14 DHH participants in locations in and around their home over a one-week period (677 recordings of 243 sound classes, avg. duration=11.5 s). To construct a dataset relevant to our evaluation, we chose participants that had recorded at least 10 classes and at least three recordings per class—resulting in nine participants (P1-P9). The samples were converted to 16Hz mono and silences greater than one second were removed. As this dataset was less balanced than in our experiment above, we could not perform similar granular experiments (*e.g.,* context-specific accuracies).

Class counts per participant, including example classes, are shown in Figure 5a. Many of the classes are highly personalized to participants' use cases (*e.g.,* flicking light switch, hearing aid whistle) and indicate that a pre-trained model would not scale well for these individuals. Moreover, existing sound datasets do not contain the requisite samples for several of these classes (*e.g.,* seatbelt alarm) to train a fully supervised model. These characteristics highlight the drawbacks of generic-model systems and reinforce the need for personalization.

| ID | Classes | Examples of recorded sounds |
|---|---|---|
| P1 | 14 | dishwasher, kettle timer, exhaust fan |
| P2 | 23 | elevator bell, bedside alarm, dumpster emptied |
| P3 | 13 | Flicking light switch, kettle, motorcycle running |
| P4 | 15 | door knock, candle lighter, fit bit alarm |
| P5 | 18 | oven timer, washer ending, garbage disposal |
| P6 | 16 | microwave beep, car engine, seatbelt alarm |
| P7 | 28 | dryer beep, bathroom faucet, oven beep |
| P8 | 18 | bathwater draining, car running, bathroom door |
| P9 | 15 | gas stove ticking, hearing aid whistle, doorbell |
| **Total** | **160** | |

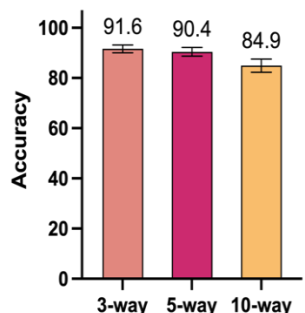

Figure 5: (a) DHH participants' recorded sound class counts with examples. Note that many of these classes are highly specific to participants' use cases (*e.g.,* flicking light switch, hearing aid whistle) and thus, require model personalization. (b) ProtoSound's average accuracy for 3-class, 5-class, and 10-class evaluations on DHH participants' recorded sounds.

For our experiment, we evaluated three settings: 3-way (3 classes), 5-way, and 10-way. We trained the model using one randomly selected recording per class for each participant (equivalent to a real-world use case) and used a clip-level prediction. See Figure 5b for results. For the 5-way setting—the most desired by DHH people—the overall accuracy was 90.4% (*SD*=4.4%). In comparison, the accuracy of the dataset's labels as rated by a hearing team member was 94.5%. Per-participant accuracies and per-class accuracies for the lowest performing participant (P8) are shown in Figure 6. Results were poor for participants P6, P7, and P8 due to two sources of errors: first, similarity among some sound classes led to confusion (*e.g.,* water draining in the bathtub *vs.* in a sink, laundry room fan *vs.* floor fan); second, some recordings did not appear to contain the labeled sounds (*e.g.,* egg cooker, car running for P8). More detailed analysis of user errors can be found in the original work [21].

We also compared performance with a supervised baseline, finding a significant increase in accuracy: for a 5-way setting, the performance difference was 19.7%, pairwise t-test yielded *t*=16.2, *p*<.001. Overall, our analysis showed ProtoSound has the potential to accommodate a wide variety of sounds from our target population.

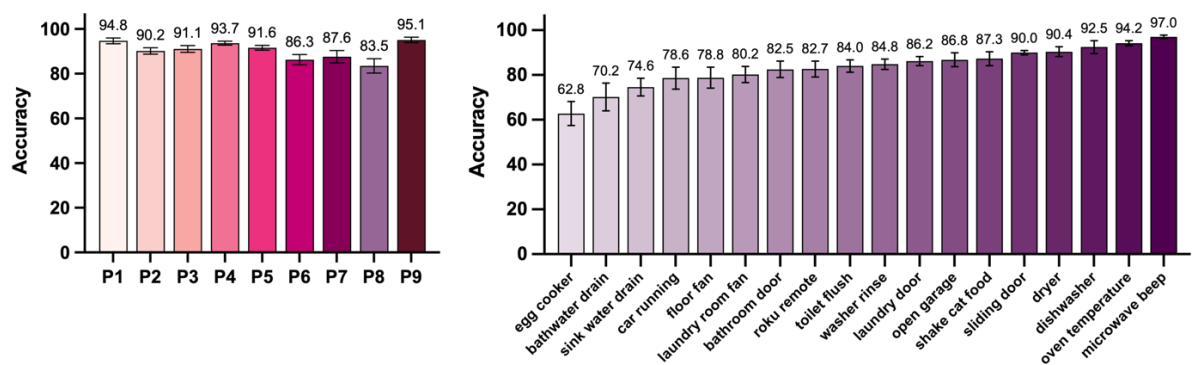

Figure 6: (a) Average accuracy per DHH participant for the 5-way setting. (b) Accuracy per-class for the lowest performing case: P8. Note that 'egg cooker' performed poorly due to user recording errors in some samples (missing sound). 'Bathwater drain' and 'sink water drain' performed poorly since they were very similar sounding and were confused with each other.

## 7 EXPERIMENT 3: FIELD EVALUATION

Our third evaluation was a field study with 19 hearing participants. While our ultimate goal is a long-term evaluation with DHH users, demonstrating real-world efficacy and improving our pipeline is important before deployments with the target population. Thus, we deployed our ProtoSound technique through an interactive mobile application and recruited hearing participants who were able to evaluate the real-world performance by reliably listening to the sounds and providing feedback on recognition. To our knowledge, our study is the first evaluation of few-shot sound recognition in the field.

### 7.1 ProtoSound Mobile Application

Our Android-based smartphone app, shown in Figure 7, contains an experimental user interface to enable hearing users to train and evaluate a sound recognition model using our ProtoSound technique. We restricted to a 5-way, 5-shot setting (5-classes and 5-samples per class) in each location to reduce our participants' recording time and effort, but ProtoSound can support any setting. Each location uses a separate sound recognition model; for training, users first enter the name of a location (*e.g.,* kitchen, restaurant) and then select the five sound classes for recording in two ways: (1) by entering the name of their own custom sound (*e.g.,* "dog bark", "doorbell" in Figure 7b), or (2) selecting a sound from a predefined list (*e.g.,* "baby cry" in Figure 7b). For each custom-defined sound, users record the five required sound samples, each of one second duration. Additionally, they can play back the recording and re-record to correct any errors. For a predefined sound, the app randomly selects five samples from ProtoSound's existing library of sounds (see Section 3.3.2). Finally, users record an ambient soundscape of the location and submit the samples for training (Figure 7c).

After training, the app saves the model, which can be used for evaluation at any time by opening the app and clicking on the evaluate tab (Figure 7d). For evaluation, the app samples the audio every four seconds (an estimation of average length of all sounds from our test set) and outputs a prediction (Figure 7e).

**Implementation.** To preserve user privacy and support offline use, the mobile app uses a *pyTorch-mobile* implementation of our ProtoSound pipeline and can run fully *on-device*. However, for the study specifically, the app interfaced with a socket.io server located at our institution, and, with each recognized sound, it displayed a binary rating form (correct, incorrect) for users to evaluate the recognition accuracy. These ratings along with other study data (user recordings, system logs) were uploaded to the server for analysis (and deleted after the analysis was complete). The entire app code is open sourced at https://github.com/makeabilitylab/ProtoSound.

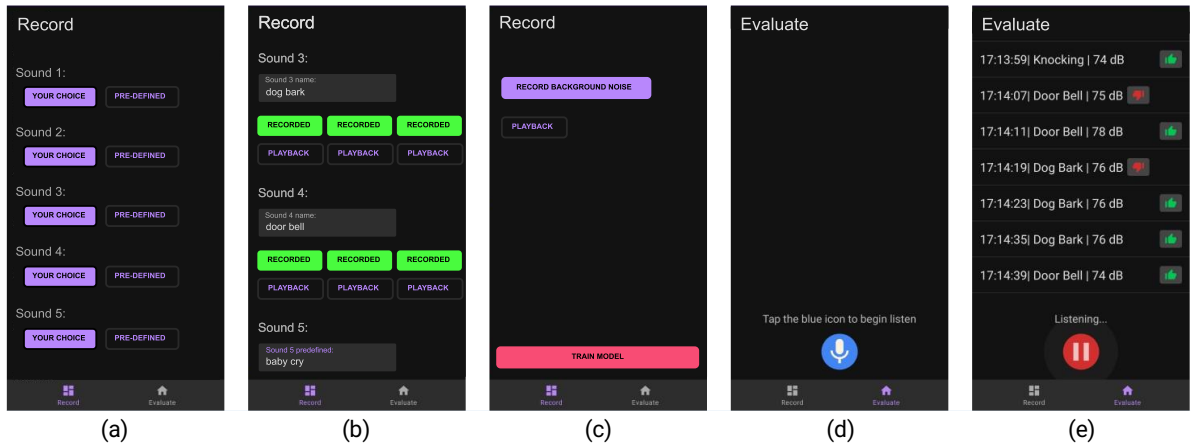

Figure 7: User interface of our ProtoSound Android application.

## 7.2 Participants

We initially recruited 20 hearing participants through various social media platforms, but one quit early due to app installation issues. The remaining 19 participants (10 women, 9 men) were on average 38.9 years old ($SD$=14.0, $range$=21-61), resided in 14 different US cities, and used an Android phone. Participants were compensated $35.

## 7.3 Procedure

The study was conducted remotely due to the COVID-19 pandemic. We emailed step-by-step instructions, with an option to ask for clarifications through email, text message, or online chat. After a link to a short demographic questionnaire, the instructions outlined how to install the app, followed by a short usage tutorial. Then, the participants chose three locations in and outside their home for app evaluation, such that: (1) each location had at-least five "naturally" occurring sounds and (2) at least one location was outdoors (*e.g.,* parking garage, park). Participants completed the recording and the evaluation tasks for five sounds in each chosen location.

For the recording task, participants could choose a predefined sound or record samples for their own sound, although, to ensure that they do not rely heavily on the existing list, they were asked to define at least one custom-defined sound in each location. During recording, the acoustic activity did not have to be naturally occurring; participants could produce the sounds themselves (*e.g.,* by deliberately knocking or turning a faucet on). This ensured high-quality recordings since it may be difficult to get an isolated single class sound sample in real-life settings. To incorporate inter-class disparity, participants were also encouraged to add everyday variations in their recordings when possible (*e.g.,* by recording different kinds of dog barks or vacuuming on different surfaces).

For the evaluation, participants rated 100 app recognitions (by thumbing up or down, see Figure 7e) in each location. Contrary to the recording task—where participants were allowed to produce the sounds themselves— the evaluations were performed in "natural" acoustic settings (*e.g.,* during meal preparation for a kitchen location, or during busy weekends in a park). Our criterion was that at least one of the five sounds had to be spontaneously occurring in a location. The app saved the state and displayed a notification when 100 ratings were complete. Participants could then end the test or optionally, rate a few more recognitions. The total evaluation time in a location was about 15 minutes, but it was not necessarily continuous—*e.g.,* participants could evaluate for 5 minutes each during breakfast, lunch, and dinner based on convenience and presence of natural acoustic activity.

After completing all the three locations, participants completed a short questionnaire to provide open-ended feedback on their experience and document examples of any sounds that were consistently correctly/incorrectly recognized or missed altogether during their evaluations.

### 7.4 Findings

We detail our mobile app's usage summary, ProtoSound's overall and context specific accuracy, sources of errors, and comparison to prior approaches.

*7.4.1 Usage summary*

All participants completed three locations, except P4 who could not evaluate an outdoor location due to quarantine requirements, resulting in a total of 56 locations (5769 sample evaluations). Figure 8a shows the locations with some example sounds. In homes, the common locations were kitchen, bedroom, and bathroom. Outdoors, participants selected parks, parking lots, and streets. Other locations included restaurants, cafes, and grocery stores. A total of 171 unique sound classes were recorded.

The total recording time per context (including set-up time, context switch time, and recording time for 25 samples of one second duration each) was on average 10.2 minutes (*SD*=3.6 minutes, *range*=6.1-18.7 minutes). This falls safely below the suggested maximum recording time from our survey (25 minutes), confirming that ProtoSound requires low-effort end-user training. The average model training time (time between submitting samples and obtaining a new model) on the users' phones was 2.4 seconds (*SD*=0.9 seconds, *range*=1.3-4.9 seconds), indicating that ProtoSound supports real-time on-device model personalization. The median total time gap between training and evaluation was 4.0 hours (*IQR*=12.8 hours, *range*=0.1-30.9 hours) and the median total evaluation span was 5.4 hours (*IQR*=18.7 hours, *range*=0.2-51.6 hours). Since the evaluations could be discontinuous, 16/56 evaluations spanned multiple days. This open-ended discontinuous evaluation allowed us to study the real-world applicability of ProtoSound.

*7.4.2 Overall and context-specific accuracies*

The average accuracy of our app across all locations was 87.4% (*SD*=6.3%). When comparing locations, the accuracy was highest in Bedrooms (*avg.*=92.6%, *SD*=3.8%) and lowest in Restaurants (*avg.*=82.2%, *SD*=7.9%), potentially due to differences in noise levels and sound types. Figure 8b shows location-specific accuracies.

Note that since participants only rated the sounds that were recognized by our app, our accuracy does not account for false negatives (*i.e.,* any unrecognized sounds). Indeed, in the feedback form, most participants (14/19) self-reported examples of sounds that were sometimes missed by our app with two participants indicating examples of "frequently" missed sounds (keys jingling and bird chirp). At the same time, participants also indicated events that were consistently recognized correctly, such as microwave beeps, door knocks, furniture sliding, door open/close, dog barks, vehicle, cart rolling, and water running, many of which are desired by DHH people [5,26].

*7.4.3 Sources of Errors*

To determine the sources of errors, we did manual analysis on user recorded samples (through listening, making visualizations), finding that about 10% of the samples contained user errors (this justifies ProtoSound's 87.4% accuracy). Specifically, we found two types of errors. First, a majority of these 10% samples did not contain the labelled sound or contained another sound of interest beyond the labeled sound, thus reducing accuracy. Second, in some cases, samples belonging into different sound categories were too similar (*e.g.,* knocking and chopping, see Figure 9) and were understandably confused with each other. This points to the need to develop better user interfaces for recording and annotating sound samples. We return to this point in the Discussion.

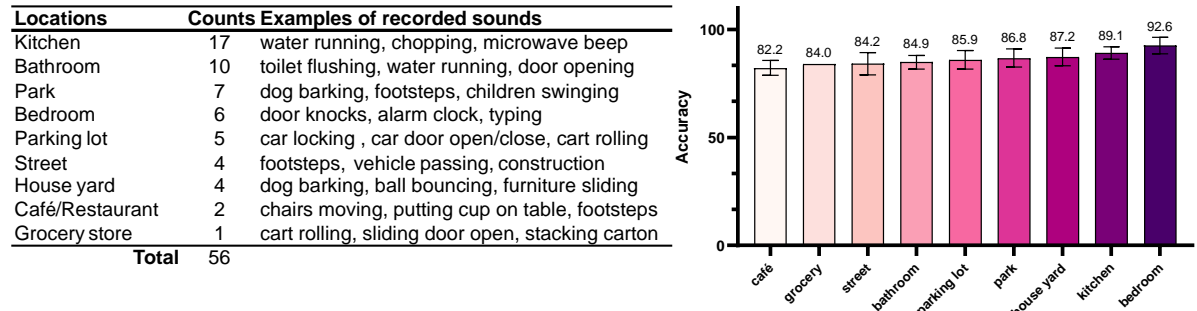

Figure 8: (a) Location for our field evaluation with the participant counts and recorded sound examples, and (b) Accuracy of our mobile app in each location. Error bars = 95% confidence intervals (no CI shown for grocery since it only had one count).

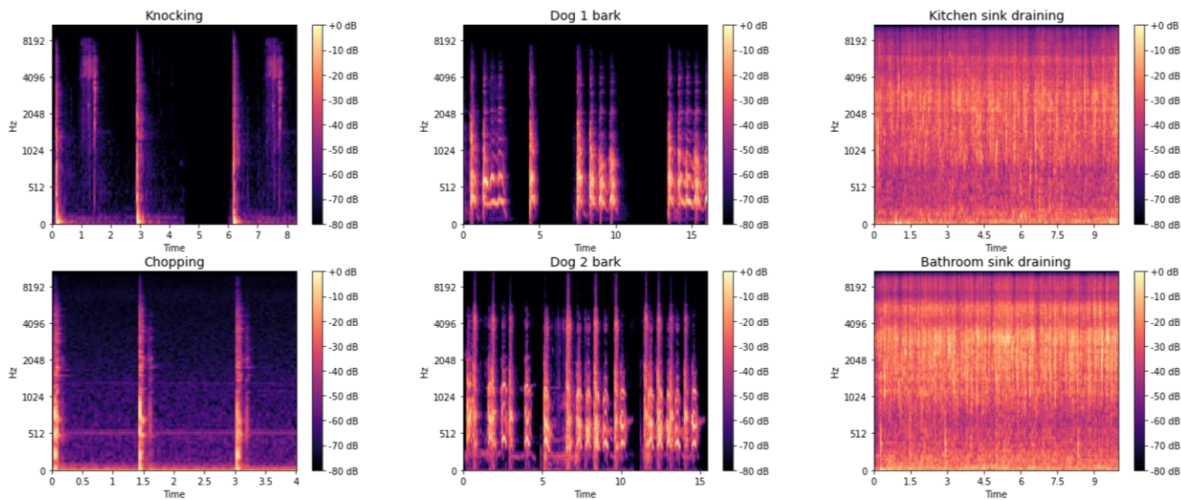

Figure 9: Mel-spectrograms of some similar sounding events that were confused with each other: knocking *vs.* chopping, barks of two different pet dogs, and kitchen *vs.* bathroom sink draining. Note the striking similarities in the spectrograms.

### 7.4.4 Effect of Pre-training and In-Situ Personalization

We also calculated the difference in performance between the sound categories that our model was pre-trained on (*e.g.,* alarm clock, door knock in train set (Table 1), total 2613 samples) and the sounds that were "unseen" by the model (*e.g.,* furniture sliding, children swinging, total 3156 samples). The average accuracy on pre-trained sound categories (91.6%, *SD*=4.7%) was higher than the new sounds (84.0%, *SD*=7.5%), suggesting that pre-training the model with the expected classes that a system may encounter in real-life will increase accuracy

For only the pre-trained sound classes, we also compared performance with state-of-the-art sound recognition systems such as *HomeSound* [25], *SoundWatch* [26], and *Ubicoustics* [33], which use a generic pre-trained VGG16 model [22]. We trained this model on our trainset and evaluated on the field samples, finding that that the accuracy was significantly lower than ProtoSound (71.3%, a difference of 20.3%) a paired t-test yielded $t_{29}$=9.1, $p$<.001. This indicates that, while allowing for new, personalized sound classes, in-situ few-shot customization also significantly improves accuracy on existing sound classes by accounting for contextual variations of sounds.

## 8 DISCUSSION

In summary, our work makes contributions in the fields of human computer interaction (HCI) and machine learning (ML) fields. Within HCI, we contribute the first personalizable sound recognition system for DHH users. Prior sound recognition systems [25,26,59] use generic pre-trained models, which do not support: (1) sounds unique to a DHH person's use case (*e.g.,* children playing), (2) variations of real-world sounds (*e.g.,* my dog vs. a generic dog), and (3) sounds with insufficient samples in existing sound datasets to train generic models (*e.g.,* footsteps). Our findings show that ProtoSound can accommodate variations in existing classes and support a variety of new, personalized classes in a diversity of contexts through low-effort end-user training.

Compared to prior ML work (*e.g.,* [58,66]) which only performed algorithmic experiments, we contribute the first deployable few-shot sound recognition system by incorporating several user-centric features in the traditional few-shot learning pipeline, such as: (1) on-the-fly training for difficult-to-produce sounds (*e.g.,* fire alarms, sirens) and (2) generalization across contexts. Our experiments show that ProtoSound significantly outperforms the best state-of-the-art few-shot baseline, yielding an accuracy improvement of 9.7% on a dataset of real-life sounds.

We also contribute findings from a large-scale survey (472 DHH users) on personalized sound recognition, insights from the first evaluation of few-shot sound recognition in the field, as well as an open-source plug-and-play system code that can be deployed on any device and a specific Android app implementation. Below we detail further implications of our work and state key limitations.

### 8.1 Graphical User Interface

**Recording and annotation interface.** Our work focused largely on building a backend pipeline for few-shot sound recognition. However, the user interface is equally important, especially for DHH users who may not be able to verify the contents of their recordings by listening to them—a challenge we observed in samples captured by DHH users. Thus, intuitive sound visualizations are essential for DHH users to better record and label their training samples. Spectrograms and waveform visualizations may be a good place to start; however, these low-dimensional features may not sufficiently represent the sample's quality [21]. Furthermore, our system expects sufficient separation among classes and enough variation among interclass samples. Failing to meet these requirements led to classification errors in our field study, which we expect will increase more when DHH users are unable to recognize auditory similarities among sounds during the training process [21]. Thus, in a parallel project, our team is researching cause-and-effect visualizations (*e.g.,* showing cluster visualization of NN-classifier) with an aim to make end-users really understand how their recorded samples may shape the model. A well-designed user-interface may even further increase our system's accuracy by improving training sample quality and reducing user errors.

**Human-in-the Loop applications.** ProtoSound is an example of Human-in-the-Loop (HITL) systems, which leverage end-user input in the model building and refining pipeline to improve performance [13,55]. However, user agency in our pipeline is limited to providing training examples. The training process itself is still a black box. Through providing more information about the training pipeline (*e.g.,* by visualizing intermediate model layers, showing measures of uncertainty in model prediction), we hope to support a greater user agency. Our pipeline can also be extended to incorporate reinforcement learning techniques [31] to gradually adapt the model by taking user feedback on the recognition output, for example, using the user interface explored in ListenLearner [68].

Such techniques could capture even greater contextual and temporal variations of real-world acoustic events beyond that is accommodated by limited samples in ProtoSound (*e.g.,* varying cries of a baby or different piano notes). However, while DHH users may be able to validate this output in a familiar location (*e.g.,* in a kitchen) with the help of visual cues, they may find it challenging to do in unpredictable contexts.

### 8.2 Algorithmic Improvements

**Acoustic event detection.** ProtoSound processes data frame-by-frame using a fixed sampling window. While this worked for our purposes, sound classes vary considerably in length—from short-lived (*e.g.,* a gunshot) to longer events (*e.g.,* thunder)—and selecting an optimal window is challenging. If the window is too small, long-term variations may not be captured. Conversely, if the window is too long, detecting boundaries between consecutive sound events is difficult. Thus, future work should explore acoustic event detection techniques (*e.g.,* sub-frame processing [28] or sequential learning [69]) to automatically segment real-life sound events.

**Collaborative learning.** Another area of exploration is federated (or collaborative) learning [70], where multiple mobile devices collaboratively learn a shared model while keeping all training data local. This technique is useful for drastically improving model performance without compromising user's data privacy, such as in the Google Keyboard (GBoard) [71] where the gesture typing technique is improved over time by averaging locally personalized models collected from billions of users [80]. Our ProtoSound pipeline is well suited for this task since the generated low-resolution class-prototypes from the model personalization process can be directly uploaded to the cloud, without compromising privacy.

**Simultaneous events.** While ProtoSound only conveys the most probable sound, our pipeline can also be modified to output multiple simultaneous events. Indeed, in past work [26], DHH users preferred the idea of showing multiple sound events in low confidence situations. However, this could easily lead to information overload, so future systems need to be carefully designed. Similarity detection techniques (*e.g.,* [4]) can be used to group multiple similar events together (*e.g.,* show "appliances" instead of "could be a microwave beep or a dishwasher"). Likewise, systems leveraging contextual information (*e.g.,* location of deployment) can ignore some detected events in a similar way a human understands that a car honk is unlikely to originate from a kitchen.

### 8.3 Socio-Cultural Implications

**Deaf culture.** While our sound recognition technology is heavily informed by DHH perspectives and past work [25,26], we do not assume it is universally desired or that it will necessarily work, as designed, for all users. Some DHH people may feel negatively towards this technology, especially those who identify as part of the Deaf culture [6,32]. However, our survey aligns with prior work [5,14] and suggests that many DHH individuals find sound recognition valuable, and even more so if it is personalized. Such a tool can be constrained to detect only a small subset of sounds (*e.g.,* a child's cry) to provide essential situational awareness while otherwise avoiding the hearing world. Still, more work is needed with the DHH population to evaluate the accessibility of our system, given diverse preferences and interests.

**Privacy.** To preserve privacy, our pipeline can run locally on devices without the need to transmit audio data to the cloud. However, uploading data has other benefits such as using it for interactively improving the classification model. Our technique can also support privacy-preserving cloud-based computation since we compute low-resolution mel-spectrograms of input data, which, while readily identifying speech, make the spoken content challenging to recover.

### 8.4 Limitations

Our work has the following primary limitations.

**Five-class setting.** First, though ProtoSound can support any number of classes, for two of our three experiments, we used a five-way (five-class) implementation since this most closely resembles what DHH people wanted in past work and in our survey. Specifically, in evaluations of generic-model sound recognition systems [25,26], DHH users enabled only 3-5 medium-to-high priority sound classes in each location to avoid being overwhelmed by notifications. Furthermore, we wanted our field study participants to spend a minimum time recording. ProtoSound's implementation is location-specific—*that is,* users train a separate model for each location, which can be switched manually, or in the future, automatically through a location-aware design (*e.g.,* [26]). Such an implementation can support, for example, 15-25 classes in a home by using a separate model for each room. Nevertheless, while few-shot learning has not yet reached a stage to support more than a few classes [48,62], ProtoSound (and its open-source implementation) can support any setting and we report on performance of different class sizes in our Experiment 2. We also encourage future work to experiment with larger class configurations while using other ways to improve performance (*e.g.,* by constraining to very specific types of sounds, or increasing the number of training samples per class).

**Survey recruitment bias.** Second, by relying on assistive technology use to identify DHH users (per institute policy), our online survey may have excluded participants who are less likely to use these technologies (*e.g.,* sign language users). We, however, reference a past survey [14] which showed that more than 75% of those who preferred sign language were interested in sound recognition support.

**Dataset constraints.** Third, we evaluated performance on a real-life dataset complied from two HCI works [25,26] instead of standard machine learning benchmarks (*e.g.,* ESC-50 [51], UrbanSound8k [56]) since these benchmarks use clean sound files and do not mimic many real-world conditions (*e.g.,* background noise, overlapping sounds, context shifts). A notable exception is Google's *AudioSet* [18], but the labelling accuracy of this publicly released dataset is very poor [81]. Nevertheless, we believe we effectively contextualized ProtoSound's performance by implementing and comparing accuracy with multiple state-of-the-art few-shot baselines on our compiled test set, which contains sound recordings from 21 real-world locations. Future work should collect and extend our experiments with larger, more varied datasets.

**Short technical evaluation.** Finally, participants in our field study used the app briefly in each location, which while demonstrating promising potential for few-shot sound recognition, does not account for a longitudinal use where a user could be moving through a range of acoustic contexts over time (*e.g.,* home to outdoors to office). While our approach should theoretically handle these contextual shifts, long-term deployments across contexts are needed to quantify the performance over a longer use period.

### 9 CONCLUSION

Sound recognition can provide important environmental, situational, and safety-related cues to people who are d/Deaf or hard of hearing (DHH). Existing sound recognition systems, however, do not support personalization to users' specific desired sounds. In this work, we presented the design and evaluation of ProtoSound, an interactive system to personalize a sound recognition engine using only a few custom recordings. ProtoSound was motivated by the prior work with DHH users, the experiences of our DHH authors, and a survey we conducted with 472 DHH participants. Evaluations on two real-life datasets and with an interactive mobile application in the field suggest that ProtoSound can support highly personalized sound categories through low end-user effort, can train the model on-device in real-time, and can handle contextual variations in a variety of real-world contexts. Beyond sound recognition, our personalization technique also has the potential to support applications in other domains such as context-aware assistants, personalized speech recognition, and home automation.

## ACKNOWLEDGMENTS

We thank Quan Dang and Hang Do for their help in deploying ProtoSound. Our work was supported by NSF Grant no: IIS-1763199.

## A  APPENDICES

### A.1 Questionnaire: DHH Survey on Personalized Sound Recognition

1. The survey will ask about medical or health topics
   - ○ Ok, got it.
   - ○ No, thanks.
2. Do you use any of the assistive technologies below on a daily or near daily basis?
   Select all answers that apply
   - ☐ Hearing aid
   - ☐ TDD, TTY, or closed captions
   - ☐ Real-time captions (*e.g.,* CART)
   - ☐ Android Live Transcribe & Sound Notification
   - ☐ Other hearing assistive devices
   - ☐ None of the above
3. How often do you use Sound Event Notifications on your Android phone (*e.g.,* dog barking, doorbell ringing, microwave beeping)?
   - ○ Multiple times a day
   - ○ Once a day
   - ○ Multiple times a week
   - ○ A few times a month
   - ○ Once per month or less frequently
   - ○ Never
4. How helpful are the Android Sound Event Notifications for you?
   - ○ 3: Extremely helpful
   - ○ 2
   - ○ 1
   - ○ 0: Neither helpful nor unhelpful
   - ○ -1
   - ○ -2
   - ○ -3: Extremely unhelpful
5. What are the reasons that you don't use Android Sound Event Notifications more often (or not at all)?
   Select all answers that apply
   - ☐ Might notify me when no sounds are occurring
   - ☐ Might miss sounds
   - ☐ Might recognize sounds incorrectly
   - ☐ Might not recognize some sounds I care about
   - ☐ Doesn't allow me to select the sounds I want
   - ☐ Might trigger too many notifications
   - ☐ Other (please specify): _________________
6. If you could define and record your own sounds for notifications, how interested will you be in doing so?
   - ○ 3: Extremely interested
   - ○ 2

- ○ 1
- ○ 0: Neutral
- ○ -1
- ○ -2
- ○ -3: Extremely uninterested

7. What is the **minimal** number of sounds that a sound recognition technology needs to support in **each context** (*e.g.,* kitchen, bedroom, restaurant) to be useful for you?
   - ○ 1-3 (safety sounds only, *e.g.,* fire alarm)
   - ○ 4-6 (plus appliance alerts, *e.g.,* kettle)
   - ○ 7-9 (plus mundane sounds, *e.g.,* vacuum)
   - ○ > 10 (practically almost all sounds)

8. If recording your own sounds could improve recognition a lot, what is the **maximum** number of sounds would you be willing to record in **each context** (*e.g.,* kitchen, bedroom)?
   ___________________

9. If recording your own sounds could improve recognition, what is the **maximum** number of minutes would you be willing to spend recording in a context (*e.g.,* kitchen, bedroom)?
   ___________________

10. How interested will you be in anonymously contributing your recorded sounds to make better sound recognition technology?
    - ○ 3: Extremely interested
    - ○ 2
    - ○ 1
    - ○ 0: Neutral
    - ○ -1
    - ○ -2
    - ○ -3: Extremely uninterested